\documentstyle[11pt,epsfig]{article}
\newcommand{\ba}{\begin{array}}
\newcommand{\ea}{\end{array}}
\newcommand{\bd}{\begin{displaymath}}
\newcommand{\ed}{\end{displaymath}}
\newcommand{\be}{\begin{equation}}
\newcommand{\ee}{\end{equation}}
\newcommand{\bea}{\begin{eqnarray}}
\newcommand{\eea}{\end{eqnarray}}

\def\etal{ {\em et al.}~}

\def\q2 {q^2}

\def\lp{\lambda^{\prime}}
\def\lps{\lambda^{\prime *}}
\def\lpp{\lambda^{\prime\prime}}
\def\lpps{\lambda^{\prime\prime * }}

\def\bapp{b_1^{\prime\prime}}
\def\bbpp{b_2^{\prime\prime}}
\def\bcp{b_3^{\prime}}
\def\bdp{b_4^{\prime}}
 \def\N10{\widetilde \chi_1^0}
                         \def\C1p{\widetilde \chi_1^+}
                         \def\C1m{\widetilde \chi_1^-}
                         \def\C1pm{\widetilde \chi_1^\pm}

\def\go{\rightarrow}
\def\beq{\begin{eqnarray}}

\def\wrp {{\cal W}_{R\!\!\!\!/}}
\def\enq{\end{eqnarray}}

\begin{document}
\begin{center}
{\Large\bf Constraining $R$-parity violating couplings
from $B\go PP $ decays using QCD improved factorization
method}\\[15mm]
{\bf Dilip Kumar Ghosh$^{a,}$\footnote{dghosh@phys.ntu.edu.tw},
Xiao-Gang He$^{a,}$\footnote{hexg@phys.ntu.edu.tw},
Bruce~H.~J.~McKellar$^{b,}$\footnote{b.mckellar@physics.unimelb.edu.au}\\ and
Jian-Qing Shi $^{a,}$\footnote{hexg1@phys.ntu.edu.tw}}\\[4mm]
{\em $^{a}$ Department of Physics, National Taiwan University\\
Taipei, TAIWAN 10617, Republic of China}\\[4mm]
{\em $^{b}$ School of Physics,\\
University of Melbourne, Parkville, Vic 3052, Australia}\\[4mm]
\end{center}
\begin{abstract}
We investigate the role of 
$R$-parity violating interaction in the
non-leptonic decays of $B$ mesons into two light mesons $B \to PP$.
The decay amplitudes are calculated using the QCD improved factorization 
method. Using the combined data on
$B$ decays from BaBar, Belle and CLEO, we obtain strong constraints on
the various products of
$R$-parity violating couplings. Many of these new constraints are stronger
than the existing bounds.
\end{abstract}

\vskip 1 true cm

\noindent

\newpage
\setcounter{footnote}{0}

\section{Introduction}
The Standard Model (SM) of particle physics is very successful in
explaining most of the elementary particle physics phenomena at the
electroweak scale.  Unfortunately, despite of its eminent success, the
SM suffers from certain drawbacks, the hierarchy problem being a major
one.  Supersymmetry (SUSY)~\cite{susy} provides an elegant solution
and, consequently has been extensively studied as a model beyond the
SM. If such a new physics exists at the electroweak scale, then it may
provide some experimental signature in the future, or even in present
data.  Looking for such new physics effects constitutes a major area
of research in high energy physics today.  There are two major ways
one can see such effects.  One is to observe their direct effects in
the high energy collider experiments when a new particle is produced
and observed through its decays.  The other way, is to look for
indirect evidence in the deviation from the SM prediction of  low
energy experimental data.

Recent results from the ongoing experiments in $B$-physics at CLEO,
BaBar, and Belle have attracted lot of attention.  Much of this
attention has been devoted to the results on nonleptonic decays of
$B$ mesons, which can be used to extract information on the CKM matrix
elements and CP violation.  The theoretical understanding of the
nonleptonic decays of $B$ mesons is an extremely demanding challenge
due to difficulties in calculating the relevant hadronic matrix
elements.  To have some idea of the magnitude of the matrix elements,
one usually uses factorization method, factorizing the four quark
operators relevant to non-leptonic $B$ decays into the products of two
currents and evaluating separately the matrix elements of the two
currents.  Recently the QCD improved factorization method for the hadronic $B$
decays has been developed. This method incorporates elements of the naive
factorization approach (as its leading term) and perturbative QCD
corrections (as subleading contributions) allowing one to compute
systematic radiative corrections to the naive factorization for the
hadronic $B$ decays~\cite{BBNS0,hn}.  In our analysis 
we will use the formalism developed in Ref.~\cite{BBNS0}.  This
QCD-improved factorization method improves the analysis on several
aspects, including among others the number of colors, the gluon virtuality, the
renormalization scale, and the scheme dependence.  The method is
expected to give a good estimate of the magnitudes of the hadronic
matrix elements in non-leptonic $B$ decays, and has been used to
calculate $B$ decays in the SM~\cite{BBNS0,BBNS1,DDG,he1} and models
beyond~\cite{he2}.

The construction of most general supersymmetric extension of the
standard model leads to baryon $(\rm B)$ and lepton $(\rm L)$ number
violating operators in the superpotential.  The simultaneous presence
of both $(\rm L)$ and $(\rm B)$ number violating operators induce
rapid proton decay which may contradict the strict experimental
bound~\cite{proton_dk}.  In order to keep the proton lifetime within
the experimental limit, one needs to impose additional symmetry beyond
the SM gauge symmetry to force the unwanted baryon and lepton number
violating interactions to vanish.  In most cases, this has been done
by imposing a discrete symmetry called $R$-parity~\cite{fayet},
defined as $R = (-1)^{3\rm B+\rm L+2\rm S}$, where, $\rm S$ is the
spin of the particle.  This symmetry not only forbids rapid proton
decay~\cite{weinberg}, it also render stable the lightest
supersymmetric particle (LSP).  However, this symmetry is {\it ad hoc}
in nature.  There are no strong theoretical arguments in support of
this discrete symmetry.  Hence, it is interesting to see the
phenomenological consequences of the breaking of $R$-parity in such a
way that either $\rm B$ or $\rm L$ number is violated both are not
simultaneously violated, thus avoiding rapid proton decays.  Extensive
studies have been done to look for the direct as well as indirect
evidence of $R$-parity violation from different processes and to put
constraints on various $R$-parity violating
couplings~[\ref{rpv_review}, \ref{Majorana}-\ref{aa}].

The main purpose of this paper is to constraint various $R$-parity
violating couplings using the data on $B \to PP$ decay channels and a
calculation based on QCD-improved factorization.  Here $P$ is one of
the S(3) flavor octet pseudoscalars. We find that using the
experimental data on the branching ratios of $B \to PP$ mode,
stringent upper bounds on the products of several $\rm L$ and $\rm B$
violating couplings can be obtained.  Many of the bounds obtained are
stronger than the existing ones.

The organization of the paper is the following.  In section (2) we
study possible four quark operators which can induce $B\to PP$ decays
with $R$-parity violating interactions.  In section (3) we calculate
$B\to \pi\pi, K \pi, K\bar K$ decay amplitudes using the QCD-improved
factorization method.  At last in section (4), we carry out numerical
analysis, present and discuss our results and finally draw our conclusions.

\section{New operators for $B\to PP$ with $R$-parity violation}

The most general superpotential of the Minimal Supersymmetric Standard Model 
(MSSM)~\cite{susy}, which describes $ SU(3)\times SU(2)\times U(1)$ 
gauge invariant, 
renormalizable and supersymmetric theory, with minimal particle content, has 
$R$-parity violating interaction terms\cite{fayet}: 
\beq
{\cal\wrp } &=& \frac{1}{2}\lambda_{[ij]k} \hat L_i\hat L_j
\hat E^c_k + \lambda^\prime_{ijk} \hat L_i \hat Q_j \hat D^c_k
+\frac{1}{2}\lambda^{\prime \prime}_{i[jk]} \hat U_i^c
 \hat D^c_j \hat D^c_k
\label{rpv1}
\enq
where, $\hat L$ and $ \hat Q $ are the $SU(2)$-doublet lepton and quark
superfields and $\hat E^c$, $\hat U^c $ and $ \hat D^c $ are the singlet
superfields, while $i,j,k$ are flavor indices. In writing the above we have
omitted gauge indices, which ensures that the $\lambda_{ijk}$ are antisymmetric
in $i$ and $j$, and the $\lambda^{\prime\prime}$ are antisymmetric in $j$ and
$k$. It is very clear that the first two terms in Equation (\ref{rpv1})
violate lepton number, while the last term violates baryon number.

The above $R$-parity violating interaction, in general
can have 27 $\lp$-type and 9 each of $\lambda$ and
$\lpp$-type of new couplings. Not all of them will induce $B\to PP$
decays to the lowest order. At the dimension six four-fermion
interaction level only
operators corresponding to the couplings $\lp_{ijk}$ and $\lpp_{ijk}$ will
lead to hadronic $B\to PP$ decays.
They are given by
\begin{eqnarray}
L_{eff}&=&\frac{\lambda^{''}_{112} \lambda^{''*}_{132}}{2  m_{ \widetilde{s}}^2}(
\bar u_\alpha \gamma_\mu R u_\alpha \bar d_\beta \gamma_\mu R b_\beta
-\bar u_\alpha \gamma_\mu R u_\beta \bar d_\beta \gamma_\mu R b_\alpha) \nonumber \\
&-&\frac{\lambda^{'}_{i11} \lambda^{'*}_{i13}}{2  m_{ \widetilde{e}_i}^2}
\bar u_\alpha \gamma_\mu L u_\beta \bar d_\beta \gamma_\mu R b_\alpha\nonumber \\
&-&\frac{\lambda^{'}_{i11} \lambda^{'*}_{i13}}{2  m_{ \widetilde{\nu}_i}^2}
\bar d_\alpha \gamma_\mu L d_\beta \bar d_\beta \gamma_\mu R b_\alpha
-\frac{\lambda^{'}_{i31} \lambda^{'*}_{i11}}{2  m_{ \widetilde{\nu}_i}^2}
\bar d_\alpha \gamma_\mu R d_\beta \bar d_\beta \gamma_\mu L b_\alpha \nonumber \\
&+&\frac{\lambda^{''}_{121} \lambda^{''*}_{131}}{2  m_{ \widetilde{d}}^2}(
 \bar u_\alpha \gamma_\mu R u_\alpha \bar s_\beta \gamma_\mu R b_\beta-\bar u_\alpha \gamma_\mu R u_\beta \bar s_\beta \gamma_\mu R b_\alpha) \nonumber \\
&+&\frac{\lambda^{''}_{i12} \lambda^{''*}_{i13}}{4  m_{ \widetilde{u}_i}^2}
( \bar d_\alpha \gamma_\mu R d_\alpha \bar s_\beta \gamma_\mu R b_\beta
-\bar d_\alpha \gamma_\mu R d_\beta \bar s_\beta \gamma_\mu R b_\alpha) \nonumber \\
&-&\frac{\lambda^{'}_{i12} \lambda^{'*}_{i13}}{2  m_{ \widetilde{e}_i}^2}
\bar u_\alpha \gamma_\mu L u_\beta \bar s_\beta \gamma_\mu R b_\alpha \nonumber \\
&-&\frac{\lambda^{'}_{i11} \lambda^{'*}_{i23}}{2  m_{ \widetilde{\nu}_i}^2}
\bar s_\alpha \gamma_\mu L d_\beta \bar d_\beta \gamma_\mu R b_\alpha
-\frac{\lambda^{'}_{i32} \lambda^{'*}_{i11}}{2  m_{ \widetilde{\nu}_i}^2}
\bar s_\alpha \gamma_\mu R d_\beta \bar d_\beta \gamma_\mu L b_\alpha \nonumber \\
&-&\frac{\lambda^{'}_{i12} \lambda^{'*}_{i13}}{2  m_{ \widetilde{\nu}_i}^2}
\bar d_\alpha \gamma_\mu L d_\beta \bar s_\beta \gamma_\mu R b_\alpha
-\frac{\lambda^{'}_{i31} \lambda^{'*}_{i21}}{2  m_{ \widetilde{\nu}_i}^2}
\bar d_\alpha \gamma_\mu R d_\beta \bar s_\beta \gamma_\mu L b_\alpha \nonumber \\
&+&\frac{\lambda^{''}_{i12} \lambda^{''*}_{i23}}{4  m_{ \widetilde{u}_i}^2}
(\bar d_\alpha \gamma_\mu R s_\alpha \bar s_\beta \gamma_\mu R b_\beta
-\bar d_\alpha \gamma_\mu R s_\beta \bar s_\beta \gamma_\mu R b_\alpha) \nonumber \\
&-&\frac{\lambda^{'}_{i22} \lambda^{'*}_{i13}}{2  m_{ \widetilde{\nu}_i}^2}
\bar d_\alpha \gamma_\mu L s_\beta \bar s_\beta \gamma_\mu R b_\alpha
-\frac{\lambda^{'}_{i31} \lambda^{'*}_{i22}}{2  m_{ \widetilde{\nu}_i}^2}
\bar d_\alpha \gamma_\mu R s_\beta \bar s_\beta \gamma_\mu L b_\alpha  \nonumber \\
&-&\frac{\lambda^{'}_{i21} \lambda^{'*}_{i23}}{2  m_{ \widetilde{\nu}_i}^2}
\bar s_\alpha \gamma_\mu L s_\beta \bar d_\beta \gamma_\mu R b_\alpha
-\frac{\lambda^{'}_{i32} \lambda^{'*}_{i12}}{2  m_{ \widetilde{\nu}_i}^2}
\bar s_\alpha \gamma_\mu R s_\beta \bar d_\beta \gamma_\mu L b_\alpha,
\end{eqnarray}
where $m_{\tilde f}$ is the sfermion mass, $L(R) = (1\mp\gamma_5)/2$, and
$\alpha,\;\beta$ are the color indices.

There are three types of four-quark operator in the above four quark 
interactions,
\begin{eqnarray}
&&O_A=(\bar p_\alpha q_\beta)_{V \pm A}(\bar r_\beta b_\alpha)_{V \mp A}.\nonumber\\
&&O_B= (\bar p_\alpha q_\alpha)_{V+A}(\bar r_\beta b_\beta)_{V+A},\nonumber\\
&&O_C=(\bar p_\alpha q_\beta)_{V+A}(\bar r_\alpha b_\beta)_{V+A},
\end{eqnarray}
where $(\bar p q)_{V\pm A} = \bar p \gamma_\mu (1\pm \gamma_5) q$.

The first operator  above 
occurs in  $\lambda'$ interactions and the last two are in
$\lpp$ interactions.
The above operators   are evaluated at the common sfermion mass scale 
$(m_{\tilde f})$ of 100 GeV.
At a scale around $\mu = m_b$, these operators will induce nonzero
matrix elements causing $B\to PP$ decays. We denote
 $c(\mu)_{A,B,C}$ the Wilson coefficients of the operators
$O_{A,B,C}$ at the scale $\mu$. Renormalization group running of these
coefficients from  $m_{\tilde f}$ to $\mu$ will modify them. For
$c(\mu)_A$, we have
\begin{eqnarray}
c(\mu)_A = \eta^{-8/\beta_0} c(m_{\tilde f})_A,
\end{eqnarray}
where $\eta = \alpha_s(m_{\tilde f})/\alpha_s({\mu})$, and $\beta_0 =
11-2f/3$ with $f$ the number of quarks with mass below $\mu$.  The
other two coefficients $c(m_{\tilde f})_B$ and $c(m_{\tilde f})_C$
when evolved down to the scale $\mu$ from the high scale $m_{\tilde
f}$ will mix and are given by
\begin{eqnarray}
&&c(\mu)_B = {1\over 2} [\eta^{2/\beta_0} (c(m_{\tilde f})_B +
c(m_{\tilde f})_C) + \eta^{-4/\beta_0}(c(m_{\tilde f})_B-c(m_{\tilde
f})_C)],\nonumber\\
&&c(\mu)_C = {1\over 2} [\eta^{2/\beta_0} (c(m_{\tilde f})_B + c(m_{\tilde f})_C)
- \eta^{-4/\beta_0}(c(m_{\tilde f})_B-c(m_{\tilde f})_C)].
\end{eqnarray}

To obtain the $B\to PP$ decay
amplitudes induced by these operators, one needs to evaluate the related
hadronic matrix elements. In the next section, we will use the QCD-improved
factorization method to carry out the analysis.

\section{$B\to PP$ decay amplitude with $R$-parity violation}

The factorization approximation has been used to provide estimates for
the decay amplitudes in $B$ decays.  Recently it has been shown that
factorization approximation in fact is supported by perturbative QCD
calculations in the heavy quark limit, and the QCD-improved
factorization method has been developed\cite{BBNS0,BBNS1}.  This new
factorization formula incorporates elements of the naive factorization
approach and introduces  corrections in the amplitudes.

In the heavy quark limit, the decay amplitude $B \to P_1 P_2$ due to
some particular operator ${\cal O}_i$ can be represented in the
form~\cite{BBNS0} \beq < P_1 P_2 \mid {\cal O}_i\mid B > = < P_2\mid
J_2\mid 0> <P_1 \mid J_1 \mid B > \bigg [ 1 + \sum_n r_n \alpha^n_{s}
+ {\cal O}(\Lambda_{QCD}/m_b) \bigg ] \enq The above result reduces to
the naive factorization if we neglect the power corrections in
$\Lambda_{QCD}/m_b$ and the radiative corrections in $\alpha_s$.  The
radiative corrections, which are dominated by the hard gluon exchange
can be computed with perturbation theory in the heavy quark limit, in
terms of the convolution of the hard scattering kernel and the light
cone distribution amplitudes of the mesons.  Then a factorization
formula for $B \to P_1 P_2 $ decay can be written as~\cite{BBNS0}:
\beq <P_1 P_2 \mid {\cal O}_i\mid B> &=&F^{B\to P_1}(0)\int_0^1 dx
T^{I}_i(x) \Phi_{P_2}(x)\\\nonumber&& + \int_0^1 dx dy d\xi
T_i^{II}(\xi, x,y)\Phi_{B}(\xi) \Phi_{P_2}(x)\Phi_{P_1}(y) \enq In the
above formula, $\Phi_{B}(\xi)$ and $\Phi_{P_i}(x)~(i =1,2)$ are the
leading twist light cone distribution amplitudes of the $B$-meson and
the light pseudoscalar mesons~\cite{lcda, ball} respectively, and the
$T^{I,II}_i$ describes the hard scattering kernel which can be
calculated in perturbative QCD~\cite{BBNS0,BBNS1,DDG}.  The diagrams
generating the hard scattering kernels $T^{I,II}$ in the SM are shown
in Figure 1.  Figures 1(a)-(d) depicts vertex
corrections, Figures~1(e) and 1(f) penguin corrections, and
Figures~1(g) and 1(h) hard spectator scattering.

Applying the QCD-improved factorization method to the operators in the
previous section, we obtain the contributions from $R$-parity violating 
interactions. Since we are considering the leading effects, we  need only 
evaluate Figures 1(a)-(d) for the vertex corrections and Figures 1(g) and 1(h)
for the hard-spectator scatterings. The penguin types are higher order 
corrections. The results are listed below:
\begin{eqnarray}
A_R(\bar B^0 \to \pi^0 \pi^0)&=&if_{\pi}(m_{B}^{2}-m_{\pi}^{2})F^{B\to\pi}_0(0)
\Big[-a^{\prime \prime}_1
\frac{\lpp_{112} \lpps_{132}}{8  m_{ \widetilde{s}}^2}
-a^{\prime}\frac{\lp_{i11} \lps_{i13}}{8  m_{ \widetilde{e}_i}^2}\nonumber \\&&
+(-R_\pi c_A+a^{\prime})(\frac{\lp_{i11} \lps_{i13}}{8  m_{ \widetilde{\nu}_i}^2}
-\frac{\lp_{i31} \lps_{i11}}{8  m_{ \widetilde{\nu}_i}^2})\Big],
\\
A_R(\bar B^0 \to \pi^+\pi^-)&=&if_{\pi}(m_{B}^{2}-m_{\pi}^{2})F^{B\to\pi}_0(0)
\Big[a^{\prime \prime}_2
\frac{\lpp_{112} \lpps_{132}}{8  m_{ \widetilde{s}}^2}
-R_\pi c_A\frac{\lp_{i11} \lps_{i13}}{8  m_{ \widetilde{e}_i}^2}\Big],\\
A_R(B^- \to \pi^- \pi^0) &=&if_{\pi}(m_{B}^{2}-m_{\pi}^{2})F^{B\to\pi}_0(0)
\Big[\frac{1}{\sqrt{2}}(-R_\pi c_A + a^{\prime})
\frac{\lp_{i11} \lps_{i13}}{8  m_{ \widetilde{e}_i}^2}\nonumber\\&&
+\frac{-1}{\sqrt{2}}(-R_\pi c_A + a^{\prime})(\frac{\lp_{i11} \lps_{i13}}
{8  m_{ \widetilde{\nu}_i}^2} -\frac{\lp_{i31} \lps_{i11}}
{8  m_{ \widetilde{\nu}_i}^2})\big],\\
A_R(\bar B^0 \to K^- \pi^+)&=&if_{K}(m_{B}^{2}-m_{\pi}^{2})F^{B\to \pi}_0(0)
\Big[a^{\prime \prime}_2
\frac{\lpp_{121} \lpps_{131}}{8  m_{\widetilde{d}}^2}
-R_K c_A \frac{\lp_{i12} \lps_{i13}}{8 m_{ \widetilde{e}_i}^2}\Big], \\
A_R(B^- \to K^- \pi^0) &=&\frac{G_{F}}{\sqrt{2}}if_{\pi}(m_{B}^{2}-m_{K}^{2})F^{B\to K}_0(0)
\Big[-\frac{1}{\sqrt{2}}a^{\prime \prime}_1
\frac{\lpp_{i12} \lpps_{i13}}{16 m_{ \widetilde{u}_i}^2}
+\nonumber \\&& \frac{1}{\sqrt{2}}(-r_{K\pi} R_K
c_A+a^{\prime})\frac{\lp_{i12}
\lps_{i13}}{8 m_{ \widetilde{e}_i}^2}
+\frac{1}{\sqrt{2}}R_\pi c_A (\frac{\lp_{i11} \lps_{i23}}
 {8 m_{ \widetilde{\nu}_i}^2}-\frac{\lp_{i32}\lps_{i11}}
{8 m_{ \widetilde{\nu}_i}^2})
+\nonumber \\&& \frac{-1}{\sqrt{2}}a^{\prime} (\frac{\lp_{i12} \lps_{i13}}
{8  m_{ \widetilde{\nu}_i}^2}-\frac{\lp_{i31} \lps_{i21}}{8 m_{ \widetilde{\nu}_i}^2})\Big],\\
A_R(B^- \to \bar K^0 \pi^-) &=&if_K(m_{B}^{2}-m_{\pi}^{2})F^{B\to\pi}_0(0)
\Big[a^{\prime \prime}_2\frac{\lpp_{i12} \lpps_{i13}}
{16 m_{ \widetilde{u}_i}^2}
+a^{\prime} (\frac{\lp_{i11} \lps_{i23}}{8 m_{ \widetilde{\nu}_i}^2}
-\frac{\lp_{i32} \lps_{i11}}{8 m_{ \widetilde{\nu}_i}^2})\nonumber \\
&& -R_K c_A(\frac{\lp_{i12} \lps_{i13}}{8  m_{ \widetilde{\nu}_i}^2}
-\frac{\lp_{i31} \lps_{i21}}{8  m_{ \widetilde{\nu}_i}^2})\Big]\\
A_R(B^0 \to K^0 \pi^0)&=&if_{\pi}(m_{B}^{2}-m_{K}^{2})F^{B\to K}_0(0)
\Big[\frac{1}{\sqrt{2}}a^{\prime \prime}_1
\frac{\lpp_{121} \lpps_{131}}{8  m_{\widetilde{d}}^2}
+\frac{1}{\sqrt{2}}a^{\prime}\frac{\lp_{i12} \lps_{i13}}{8 m_{ \widetilde{e}_i}^2}\nonumber\\&&
+\frac{-1}{\sqrt{2}}(-R_\pi c_A+r_{K\pi} a^{\prime})(\frac{\lp_{i11} \lps_{i23}}
{8 m_{ \widetilde{\nu}_i}^2}-\frac{\lp_{i32} \lps_{i11}}
{8 m_{ \widetilde{\nu}_i}^2})\nonumber\\&&
+\frac{-1}{\sqrt{2}}(-r_{K\pi} R_K c_A+ a^{\prime})(\frac{\lp_{i12} \lps_{i13}}
{8  m_{ \widetilde{\nu}_i}^2}-\frac{\lp_{i31} \lps_{i21}}
{8  m_{ \widetilde{\nu}_i}^2})\Big]\\
A_R(B^- \to K^- K^0)&=&A_R(\bar B^0 \to K^0 \bar K^0) \nonumber \\
&=&if_{K}(m_{B}^{2}-m_{K}^{2})F^{B\to K}_0(0)
\Big[ a^{\prime \prime}_1 \frac{\lpp_{i12} \lpps_{i23}}
{16 m_{ \widetilde{u}_i}^2}
+a^{\prime}( \frac{\lp_{i22} \lps_{i13}}{8 m_{ \widetilde{\nu}_i}^2}
-\frac{\lp_{i31} \lps_{i22}}{8 m_{ \widetilde{\nu}_i}^2})\nonumber \\
&&-R_K c_A (\frac{\lp_{i21} \lps_{i23}}{8 m_{ \widetilde{\nu}_i}^2}
- \frac{\lp_{i32} \lps_{i12}}{8 m_{ \widetilde{\nu}_i}^2})\Big],\nonumber\\
A_R(\bar B^0 \to K^- K^+)&=&0,
\end{eqnarray}
where $f_i$, $F_0^i$ are decay constant and form factors, respectively.
The parameters $R_i$ and $r_i$ are given by,
\begin{eqnarray}
&&R_\pi = \frac{2m_{\pi}^2}{\bar m_b(\mu)(\bar m_u(\mu)+\bar m_d(\mu))},\\
&&R_K = \frac{2m_{K}^2}{\bar m_b(\mu)(\bar m_q(\mu)+\bar m_s(\mu))},\\
&&r_{K\pi }= \frac{f_K F_0^{B\to \pi}(0)(m_B^2-m_\pi^2)}{f_\pi 
F_0^{B\to K} (0)(m_B^2-m_K^2)}
\end{eqnarray}
with $q=u$ for charged kaon and $q=d$ for neutral kaon.
The parameters $a^{\prime \prime}_i$ and $a^\prime $ are defined as
\begin{eqnarray}
&&a_1^{\prime \prime} = c(\mu)_B +{c(\mu)_C\over N_c}\left[1 
+ {C_F \alpha_s \over 4\pi} V_{P_2} \right]
+\frac{c(\mu)_C}{N_c}\frac{C_F \pi \alpha_s}{N_c}H_{P_2 P_1},\\
&&a_2^{\prime \prime} = c(\mu)_C +{c(\mu)_B\over N_c}\left[1 + 
{C_F \alpha_s \over 4\pi} V_{P_2} \right] 
+\frac{c(\mu)_B}{N_c}\frac{C_F \pi \alpha_s}{N_c}H_{P_2 P_1},\\
&&a^{\prime} ={c(\mu)_A\over N_c}\left[1 - {C_F \alpha_s \over 4\pi} 
V_{P_2}^{\prime} \right]
-\frac{c(\mu)_A}{N_c}\frac{C_F \pi \alpha_s}{N_c}
H_{P_2 P_1}^{\prime},\\
&&c_A=c(\mu)_A
\end{eqnarray}
Here $N_c=3$ is the number of colors, $C_F = (N^2_c-1)/2N_c$, 
$P_1$ is the final state meson absorbing
the light spectator quark from $B$-meson, while $P_2$ is another final state
light meson which is composed of the quarks produced from the weak decay 
of $b$ quark.

$V_P^{(')}$ comes from vertex corrections 
($P=\pi,K$)(first four diagrams of Figure~1) is
give by
\begin{eqnarray}
V_P &=& 12 \ln{m_b \over \mu} - 18 + \int^1_0 dx g(x) \phi_P(x),\\
V_P^{\prime} &=& 12 \ln{m_b \over \mu} - 6 + \int^1_0 dx g(1-x) \phi_P(x),\\
g(x)&=&  3({1-2x\over 1-x} \ln x -i\pi)+
\Big[ 2 Li_2(x)-\ln^2 x+\frac{2\ln x}{1-x}
-(3+2i\pi)-(x\leftrightarrow 1-x)\Big].
\end{eqnarray}

The contributions from the hard spectator scattering as shown in the last
two diagrams in Figure~2 give leading twist and chirally-enhanced twist-3
contributions to $T^{II}_i$  parametrized by $H_{P_2 P_1}^{(\prime)}$. 
The detailed expressions can be found in~\cite{BBNS1}.

From the above, the amplitude $A_R(\bar B^0\to K^-K^+)$ is zero
without annihilation contributions (shown in Figure 2). When
annihilation contributions are included, this amplitude becomes
nonzero and  additional contributions to other decay
amplitudes are generated.  We list these contributions in the following:
\begin{eqnarray}
&&A_R^{ann}(\bar B^0 \to \pi^0 \pi^0)= A_R^{ann}(\bar B^0 \to \pi^+ \pi^-)
\nonumber \\
&&= i f_B f_\pi^2 \Big[
\bbpp \frac{\lpp_{112} \lpps_{132}}{8 m_{ \widetilde{s}}^2}-\bdp
\frac{\lp_{i11} \lps_{i13}}{8 m_{ \widetilde{e}}^2}-(\bcp+\bdp)(
\frac{\lp_{i11} \lps_{i13}}{8 m_{ \widetilde{\nu}_i}^2}-
\frac{\lp_{i31} \lps_{i11}}{8 m_{ \widetilde{\nu}_i}^2})\Big],\\
&&A_R^{ann}( B^- \to \pi^- \pi^0)=0,\\
&&A_R^{ann}(\bar B^0 \to K^- \pi^+)=-\sqrt{2} A_R^{ann}(\bar B^0 \to
\bar K^0 \pi^0)\nonumber \\
&&=i f_B f_\pi f_K \Big[\bapp\frac{\lpp_{i12} \lpps_{i13}}{16 m_{ \widetilde{u}_i}^2}-\bdp (
\frac{\lp_{i11} \lps_{i23}}{8 m_{ \widetilde{\nu}_i}^2}-
\frac{\lp_{i32} \lps_{i11}}{8 m_{ \widetilde{\nu}_i}^2})-\bcp (
\frac{\lp_{i12} \lps_{i13}}{8 m_{ \widetilde{\nu}_i}^2}-
\frac{\lp_{i31} \lps_{i21}}{8 m_{ \widetilde{\nu}_i}^2})\Big],\\
&&A_R^{ann}( B^- \to K^0 \pi^-)=\sqrt{2} A_R^{ann}( B^- \to K^- \pi^0)\nonumber \\
&&=i f_B f_\pi f_K \Big[\bapp\frac{\lpp_{121} \lpps_{131}}{8 m_{ \widetilde{d}}^2}-\bcp
\frac{\lp_{i12} \lps_{i13}}{8 m_{ \widetilde{e_i}}^2}\Big],\\
&&A_R^{ann}(\bar B^0 \to \bar K^0 K^0)=i f_B f_K^2 \Big[
-(\bcp+\bdp)(\frac{\lp_{i11} \lps_{i13}}{8 m_{ \widetilde{\nu}_i}^2}-
\frac{\lp_{i31} \lps_{i11}}{8 m_{ \widetilde{\nu}_i}^2})+\bapp
\frac{\lpp_{i12} \lpps_{i13}}{16 m_{ \widetilde{u}_i}^2}\nonumber\\&&-\bcp (
\frac{\lp_{i22} \lps_{i13}}{8 m_{ \widetilde{\nu}_i}^2}-
\frac{\lp_{i31} \lps_{i22}}{8 m_{ \widetilde{\nu}_i}^2})-\bdp (
\frac{\lp_{i21} \lps_{i23}}{8 m_{ \widetilde{\nu}_i}^2}-
\frac{\lp_{i32} \lps_{i12}}{8 m_{ \widetilde{\nu}_i}^2})\Big],\\
&&A_R^{ann}(\bar B^0 \to K^+ K^-)=i f_B f_K^2 \Big[
\bbpp \frac{\lpp_{112} \lpps_{132}}{8 m_{ \widetilde{s}}^2}-\bdp
\frac{\lp_{i11} \lps_{i13}}{8 m_{ \widetilde{e}}^2}+\bbpp
\frac{\lpp_{i12} \lpps_{i13}}{16 m_{ \widetilde{u}_i}^2}\nonumber\\&&-\bcp (
\frac{\lp_{i22} \lps_{i13}}{8 m_{ \widetilde{\nu}_i}^2}-
\frac{\lp_{i31} \lps_{i22}}{8 m_{ \widetilde{\nu}_i}^2})-\bdp (
\frac{\lp_{i21} \lps_{i23}}{8 m_{ \widetilde{\nu}_i}^2}-
\frac{\lp_{i32} \lps_{i12}}{8 m_{ \widetilde{\nu}_i}^2})\Big],\\
&&A_R^{ann}(B^- \to K^- K^0)=i f_B f_K^2 \Big[
\bapp \frac{\lpp_{112} \lpps_{132}}{8 m_{ \widetilde{s}}^2}-\bcp
\frac{\lp_{i11} \lps_{i13}}{8 m_{ \widetilde{e}}^2}\Big].
\end{eqnarray}
where
\begin{eqnarray}
&&b_1^{\prime \prime}=\frac{C_F}{N_c^2}c(\mu)_A A_1^i,\;\;\;\;
b_2^{\prime \prime}=\frac{C_F}{N_c^2}c(\mu)_B A_1^i,\nonumber \\
&&b_3^{\prime}=\frac{C_F}{N_c} c(\mu)_C A_3^f,\;\;\;\;
b_4^{\prime}=\frac{C_F}{N_c^2}c(\mu)_C A_2^i.
\end{eqnarray}
The parameters $A_i$ are given by
\begin{eqnarray}
&&A_1^i=A_2^i=\pi \alpha_s \left[ 18
\left(X_A-4+\frac{\pi^2}{3} \right)+2r_\chi^2 X_A^2 \right],\nonumber\\
&&A_3^f=12\pi \alpha_s r_\chi^2 (2X_A^2-X_A),
\end{eqnarray}
where $r_\chi\approx R_\pi$, $X_A=\int^1_0 dy/y$ parameterizes the divergent
endpoint integrals, we take the value 
$X_A=\ln(m_B/ \Lambda_h)$ with 
$\Lambda_h=0.5$~GeV being the typical hadronic scale.

\section{Results and Discussion}
In this section we discuss the constraints on the product of a pair
of $R$-parity violating couplings.
With $R$-parity violating interactions, the total $B\to PP$ decay amplitudes 
contain 
the SM part of the decay amplitudes $A_{SM}$ which have been obtained 
in Ref.~\cite{BBNS1} plus the R-parity violating amplitudes obtained in the 
previous section. Comparing the branching ratios obtained theoretically with
known experimental data, constraints on the $R$-parity violating interactions
can be obtained.

As we have shown in the previous section that  several
combinations of $R$-parity violating couplings can contribute to
a particular charmless hadronic $B\to PP$ decay. 
In our numerical analysis, we assume that
only one pair of $R$-parity violating couplings are nonzero at a time. 
This restriction
may seem to be unnatural, however, it is an useful approach that allows
one a quantitative feeling of the various experimental constraints.

For the numerical computations we use the averaged value of the
experimental input as shown in Table~\ref{expdata}.
We average the data from BaBar, Belle, CLEO B factories as their 
results are uncorrelated. If the experimental branching ratio has only upper 
limits, we select the most stringent one. 
It is clear from the Table~\ref{expdata}
that for $B \to \pi^-\pi^0, B\to \pi^0\pi^0$ and $B\to KK $ modes have only
the $90\%$ C.L. upper limit while others have results at $95\%$ C.L.

In our numerical calculation we use
$m_b(m_b)=4.2~{\rm~GeV}$ for the b quark mass. For the other lighter quark 
masses we take their central values $m_c(m_b) = 1.3\pm 0.2$~GeV, 
$m_s(2~\rm GeV) = 110.0\pm 0.25$~MeV, $ (m_u+m_d)(2\rm GeV)= 
9.1\pm 2.1 $ MeV. For the decay constants~\cite{proton_dk} and 
form factors\cite{form_fac}, we use $f_\pi = 0.131~{\rm~GeV}$, 
$f_{K} = 0.160~{\rm~GeV}$, $f_B = 0.180~{\rm~GeV}$, $F^{B\to \pi} = 0.28$, 
$r_{\pi K}\simeq \frac{F^{B\to K} f_\pi}{F^{B\to \pi} f_K}=0.9$.
For the KM matrix elements, we fix $\lambda = 0.2196$,
and take the central values of $\mid V_{cb}\mid = 0.0402\pm 0.0019 $
$\mid\frac{V_{ub}}{V_{cb}}\mid = 0.090\pm 0.025$~\cite{proton_dk}, 
but allow the CP violating phase $\gamma$ to vary
from 0 to $2\pi$. In the SM, the CP violating phase $\gamma$ is well
constrained~\cite{he3}. However, with $R$-parity violation this constraint may
be relaxed. To accommodate this we vary $\gamma$ from 0 to $2\pi$ to obtain
conservative limits on $R$-parity violating couplings.

We also need to know various light cone distribution amplitudes.
The leading-twist light cone distribution amplitudes of the light pseudoscalar
mesons~\cite{ball} can be expanded in Gegenbauer polynomials. We truncate 
this expansion at
 $n=2$.
\begin{eqnarray}
\phi_P(x,\mu) = 6x(1-x)\left[ 1+\alpha_1^P(\mu)C_1^{(3/2)}(2x-1)
+\alpha_2^P(\mu) C_2^{(3/2)}(2x-1)\right],
\end{eqnarray}
where $C_1^{(3/2)}(u)=3u$ and $C_2^{(3/2)}(u)= \frac{3}{2}(5u^2-1)$.
The distribution amplitude parameters $\alpha_{1,2}^P$ for $P=K,\pi$ are:
$\alpha_1^K=0.3$, $\alpha_2^K=0.1$, $\alpha_1^\pi=0$ and $\alpha_2^\pi
=0.1$~\cite{BBNS1}.

The hard spectator contributions to the coefficients $a^{\prime\prime}_{1,2}$,
and  $a^{\prime}$ are parametrized in terms of a single (complex) quantity
$H_{P_2 P_1}^{(')}$, which 
suffers from large theoretical uncertainties related to
the regularization of the divergent endpoint integral.
Following ~\cite{BBNS1} we use
$H_{\pi K} = H_{KK} = 0.99$ at the scale $\mu = m_b$ and
\beq
H^\prime_{P_2P_1} = H_{P_2P_1}, H_{\pi\pi}=H_{K\pi} = r_{\pi K} H_{\pi K}.
\enq

Here we should like to discuss the numerical changes of the
coefficients $a^{\prime\prime}_{1,2}$, and $a^{\prime}$ due to our use
of QCD improved factorization calculations instead of the naive
factorization approximation.  For example, in the $B \to \pi \pi$
decay mode using the naive factorization scheme the coefficients
$a^{\prime\prime}_{1}$, and $a^{\prime}$ have values 0.939455 and
0.661932 respectively.  Using the QCD improved factorization method we
get $1.0164+0.105414~i$ and $0.581234+0.148545 i$ for
$a^{\prime\prime}_{1}$ and $a^{\prime}$ respectively.  For the
coefficient $a^{\prime\prime}_{1}$ we get an enhancement of the order
$10\%$ in the real part and a new imaginary contribution, while for
the coefficient $a^{\prime}$ we have a negative contribution of the
order $12\%$ in the real part but a new imaginary contribution as
before.  The QCD improved factorization gets contributions from vertex
corrections, hard scattering and from weak annihilation.  The
contribution from annihilation diagrams is small compared with the
current-current contributions.  It is less than $5 \%$ in decays
involving $a^{\prime \prime}_{1,2}$ while for decays only involving
$a'$ the annihilation contributions can be as large as 20\%.

The results are shown in Tables~\ref{rpv_b} and ~\ref{rpv_l}.  In the
Table~\ref{rpv_b} we display limits on different pair of $\lpp$-type
couplings at $95\%$~C.L. For the published experimental data with
$90\%$C.L. it is hard to get the upper limit corresponding to
$95\%$C.L., so we denote those bounds by $(*)$ in the
Table~\ref{rpv_b}.  In Table~\ref{rpv_l} we show the bounds for the
product of $\lp$-type of couplings.  The bounds are obtained assuming
100~GeV common sfermion masses, so for other values of the sfermino
mass, the bounds on the couplings can be easily obtained by scaling
them by $m^2_{\tilde f }/(100\mbox{GeV})^2$.

We note that for several decays $R$-parity violation contributes only
through the annihilation contribution.  There are possible large
uncertainties in the evaluation of the weak annihilation contribution
to charmless $B$ decays due to the fact that they are power suppressed
in the heavy quark limit and the weak annihilation effects exhibit
endpoint singularities.  We therefore denote the modes which only
receive annihilation contributions with $(\dagger)$ to remind that
there may be large uncertainties there.

From the Table~\ref{rpv_b} and Table~\ref{rpv_l} we choose the pairs
of $R$-parity violating couplings with most stringent bounds and we
display them in Table~\ref{rpv_tot} with the existing limits on such
product of couplings.  Among the existing limits, there are several
cases, where there is no direct limit on the products of the
couplings, in that case we take the products of their individual
bounds from Ref.\cite{rpv_review}.  As it can be seen in
~\cite{rpv_review} that the bounds on the first two generation
individual couplings are much stronger than the third generation.  For
this reason, in most of the cases, the product of either or both first
two generation individual couplings are much stronger than our
prediction.  Furthermore, we find that bounds on the pair of couplings
$\lpp_{112}\lpp_{113}$, and $\lpp_{112}\lpp_{123}$ obtained from $n-\bar
n$ oscillation and double nucleon decays, $\lp_{i31}\lp_{i22}$,
$\lp_{232}\lp_{211}$ and $\lp_{332}\lp_{311}$, obtained from $\Delta
m_K$ are stronger than ours. It is, however, interesting to note that, more
than half of the bounds obtained in this paper are better than the
existing ones.  In addition to that, the bounds on
$\lpp_{212}\lpp_{213}$, $\lpp_{312}\lpp_{313}$,
$\lpp_{212}\lpp_{223}$, and $\lpp_{312}\lpp_{323}$ couplings are
completely new, in a sense that there were no previous bounds on these
pairs of couplings from any experimental data, they were from
perturbative unitarity.  In two of these cases
($\lambda_{212}^{''}\lambda_{213}^{''}$ and
$\lambda_{312}^{''}\lambda_{313}^{''}$) we have improved the
existing bound by over two orders of magnitude.

Before we go conclude we would like mention  the possible
theoretical uncertainties in the bounds arising from several input
parameters.  As described before, we have obtained all of our bounds by
using the central values for the relevant parameters, the CKM matrix
elements, the quark masses, the form factors.  We have also
parametrized the hard spectator contribution by some default number. 
Some of these parameters may change up to $20\%$, and the uncertainty in
the hard spectator contribution may be even larger.  Even allowing for
this, we do not expect our result to change more than factor of two.

In this paper we have studied $B \to \pi\pi$, $B\to \pi K$ and $B\to
K\bar K$ decays based on the QCD improved factorization approach in
the presence of $R$-parity violating couplings.  Comparing our
calculated branching ratios $B\to PP$ with the experimental data, we
have obtained bounds on the products of two $R$-parity violating
couplings.  We found that most of the bounds we obtained on 
combinations of the $\lp$ and $\lpp$ type of couplings are stronger than
the existing limits.  Rare hadronic B decays can provide important
information about $R$-parity violating interactions.

\vskip 1 true cm
\begin{center}
{\bf Acknowledgements}
\end{center}
This work was partially supported by the Australian Research Council,
the National Science Council of R.O.C. under grant numbers NSC90-2112-M002-031
and NSC90-2811-M002-054, and by the Ministry of Education Academic Excellence 
Project 89-N-FA01-1-4-3.

\newpage
\def\pr#1, #2 #3 { {\em Phys. Rev.}         {\bf #1},  #2 (19#3)}
\def\prd#1, #2 #3{ {\em Phys. Rev.}        {D \bf #1}, #2 (19#3)}
\def\pprd#1, #2 #3{ {\em Phys. Rev.}       {D \bf #1}, #2 (20#3)}
\def\prl#1, #2 #3{ {\em Phys. Rev. Lett.}   {\bf #1},  #2 (19#3)}
\def\pprl#1, #2 #3{ {\em Phys. Rev. Lett.}   {\bf #1},  #2 (20#3)}
\def\plb#1, #2 #3{ {\em Phys. Lett.}        {\bf B#1}, #2 (19#3)}
\def\pplb#1, #2 #3{ {\em Phys. Lett.}        {\bf B#1}, #2 (20#3)}
\def\npb#1, #2 #3{ {\em Nucl. Phys.}        {\bf B#1}, #2 (19#3)}
\def\pnpb#1, #2 #3{ {\em Nucl. Phys.}        {\bf B#1}, #2 (20#3)}
\def\prp#1, #2 #3{ {\em Phys. Rep.}        {\bf #1},  #2 (19#3)}
\def\zpc#1, #2 #3{ {\em Z. Phys.}           {\bf C#1}, #2 (19#3)}
\def\epj#1, #2 #3{ {\em Eur. Phys. J.}      {\bf C#1}, #2 (19#3)}
\def\mpl#1, #2 #3{ {\em Mod. Phys. Lett.}   {\bf A#1}, #2 (19#3)}
\def\ijmp#1, #2 #3{{\em Int. J. Mod. Phys.} {\bf A#1}, #2 (19#3)}
\def\ptp#1, #2 #3{ {\em Prog. Theor. Phys.} {\bf #1},  #2 (19#3)}
\def\jhep#1, #2 #3{ {\em J. High Energy Phys.} {\bf #1}, #2 (19#3)}
\def\pjhep#1, #2 #3{ {\em J. High Energy Phys.} {\bf #1}, #2 (20#3)}
\def\epj#1, #2 #3{ {\em Eur. Phys. J.}        {\bf C#1}, #2 (19#3)}
\def\eepj#1, #2 #3{ {\em Eur. Phys. J.}        {\bf C#1}, #2 (20#3)}
 
\newpage
\begin{table}[ht]
\footnotesize
\begin{center}
\begin{tabular}{|c|c|c|c|c|} \hline
Decay Mode & CLEO & Belle & BaBar & Average \\
\hline
$B^0 \to \pi^+ \pi^- $ & $4.3^{+1.6}_{-1.4}\pm 0.5 $ & $ 5.6^{+2.3}_{-2.0}\pm 0.4 $ & $ 4.1 \pm 1.0 \pm 0.7 $ & $ 4.4\pm 0.9 $ \\
\hline
$ B^- \to \pi^- \pi^0 $ & $ < 12.7 ~(90\% {\rm~C.L.})$ & $ < 13.4 ~(90\% {\rm~C.L.})$ &  $ < 9.6~(90\% {\rm~C.L.})$ &  $ < 9.6~(90\% {\rm~C.L.}) $ \\
\hline
$B^0 \to \pi^0 \pi^0 $ &  $ < 5.7~(90\% {\rm~C.L.})$ & - & - &  $ < 5.7 ~(90\% {\rm~C.L.})$ \\
\hline
$B^0 \to K^+ \pi^- $ & $ 17.2^{+2.5}_{-2.4} \pm 1.2 $ & $ 19.3^{+3.4+1.5}_{-3.2-0.6} $ & $ 16.7\pm 1.6 \pm 1.3 $ & $17.3 \pm 1.5 $ \\
\hline
$ B^- \to K^- \pi^0 $ & $ 11.6^{+3.0+1.4}_{-2.7-1.3}$ & $16.3^{+3.5+1.6}_{-3.3-1.8}$ & $ 10.8^{+2.1}_{-1.9}\pm 1.0 $ & $12.1 \pm 1.7 $ \\
\hline
$B^- \to  K^0 \pi^- $ & $ 18.2^{+4.6}_{-4.0}\pm 1.6 $ & $ 13.7^{+5.7+1.9}_{-4.8-1.8}$ & $ 18.2^{+3.3}_{-3.0} \pm 2.0 $ & $ 17.3 \pm 2.7 $ \\
\hline
$B^0 \to K^0 \pi^0$ & $ 14.6^{+5.9+2.4}_{-5.1-3.3}$ &
$ 16.0^{+7.2+2.5}_{-5.9-2.7}$ & $ 8.2^{3.1}_{-2.7}\pm 1.2 $ & $ 10.4\pm 2.7 $\\
\hline
$B^0 \to K^+ K^-$ &  $ < 1.9~(90\% {\rm~C.L.})$ &  $ < 2.7~(90\% {\rm~C.L.})$ &
  $ < 2.5~(90\% {\rm~C.L.})$ &  $ < 1.9~(90\% {\rm~C.L.})$ \\
\hline
$B^- \to K^- K^0 $ &  $ < 5.1~(90\% {\rm~C.L.})$ &  $ < 5.0~(90\% {\rm~C.L.})$ & $ < 2.4~(90\% {\rm~C.L.})$ & $ < 2.4~(90\% {\rm~C.L.})$ \\
\hline
$ B^0 \to K^0 \bar K^0 $ &  $ < 17.0~(90\% {\rm~C.L.})$ & - &  $ < 7.3~(90\% {\rm~C.L.})$ &  $ < 7.3~(90\% {\rm~C.L.})$  \\
\hline
\end{tabular}
\end{center}
\caption{Experimental results for the CP-averaged $B\to \pi\pi$, $B\to \pi K$
and $B \to KK $ branching ratios in units of $10^{-6}$\cite{Bexp}.}
\label{expdata}
\end{table}
\begin{table}[ht]
\footnotesize
\begin{center}
\begin{tabular}{|c|c|c|c|c|} \hline
&$B\to \pi\pi$ & $B\to K\pi$ & $B\to \pi K$ &$B\to K \bar K$\\ \hline
$c(m_b)_A$&\multicolumn{4}{|c|}{$1.986$}\\ \hline
$c(m_b)_B$&\multicolumn{4}{|c|}{$1.409$}\\ \hline
$c(m_b)_C$&\multicolumn{4}{|c|}{$-1.409$}\\ \hline
$a^{\prime \prime}_1$&$1.016+0.105i$&$1.002+0.105i$&$0.998+0.137i$&$1.020+0.074i$\\ \hline
$a^{\prime \prime}_2$&$-1.016-0.105i$&$-1.002-0.105i$&$-0.998-0.137i$&$-1.020-0.074i$\\ \hline
$a^{\prime}$&$0.581+0.149i$&$0.561+0.149i$&$0.607+0.104i$&$0.535+0.193i$\\ \hline
$r_A b^{\prime \prime}_1$&\multicolumn{4}{|c|}{$0.016$}\\ \hline
$r_A b^{\prime \prime}_2$&\multicolumn{4}{|c|}{$-0.016$}\\ \hline
$r_A b^{\prime}_3$&\multicolumn{4}{|c|}{$0.141$}\\ \hline
$r_A b^{\prime}_4$&\multicolumn{4}{|c|}{$0.022$}\\ \hline
\end{tabular}
\end{center}
\caption{Wilson coefficients, and related parameters due
to $R$-parity violating interactions in different decays, $\mu=m_b$,
assuming $100$~GeV sfermion mass. The first final state meson is
made of spectator quark, while the second final state meson is formed of 
quark pairs originating from $b$ quark weak decay vertex.
In the above $r_A = f_Bf_\pi/(m^2_B F^{B\to \pi}_0(0))$.}
\label{rpvcoeff}
\end{table}
\newpage
\begin{table}[ht]
\footnotesize
\begin{center}
\begin{tabular}{|c|c|c|} \hline
Couplings & Bound & {Process} \\
\hline
 & ${3.69\times 10^{-3}}^{*}$& $\bar B^0 \to \pi^0 \pi^0 $\\
\cline{2-3}
      &  $ 4.59\times 10^{-3} $ & $\bar B^0 \to \pi^+ \pi^- $\\
\cline{2-3}
$\mid \lpp_{112}\lpp_{132}\mid $ & ${3.30\times 10^{-3}}^{*}$& $\bar B^0 \to K^- K^0 $\\
\cline{2-3}
 & ${4.46\times 10^{-3}}^{*}$& $\bar B^0 \to K^0 \bar K^0 $\\
\cline{2-3}
      &  $ 3.46\times 10^{-2*\dag} $ & $\bar B^0 \to K^+ K^- $\\
\hline
       & $ 5.03\times 10^{-1} $ & $ \bar B^0 \to K^- \pi^+ $ \\
\cline{2-3}

$\mid \lpp_{121}\lpp_{131}\mid $ & $ 1.58 \times 10^{-2} $ &  $ B^- \to K^- \pi^0$\\
\cline{2-3}
       & $ 1.03 \times 10^{-2} $ & $ B^- \to \bar K^0 \pi^-$\\
\cline{2-3}
        & $ 6.80 \times 10^{-3} $ & $ \bar B^0 \to \bar K^0 \pi^0$\\
\hline
       & $ 7.45\times 10^{-1\dag} $ & $ \bar B^0 \to K^- \pi^+ $ \\
\cline{2-3}
$ \mid \lpp_{i12}\lpp_{i13}\mid $ & $ 1.58 \times 10^{-2} $ &  $ B^- \to K^- \pi^0$\\
\cline{2-3}
 $(i\neq 1)$       & $ 1.03 \times 10^{-2} $ & $ B^- \to \bar K^0 \pi^-$\\
\cline{2-3}
        & $ 7.55 \times 10^{-1\dag} $ & $ \bar B^0 \to \bar K^0 \pi^0$\\
\hline
        & $ {3.19 \times 10^{-3}}^{*} $ & $ B^- \to K^- K^0$\\
\cline{2-3}
$ \mid \lpp_{i12}\lpp_{i23}\mid $         & $ {4.46 \times 10^{-3}}^{*} $ &  $ \bar B^0 \to K^0 \bar K^0$\\
\cline{2-3}
   $(i\neq 1)$      & $ {1.04 \times 10^{-1}}^{*\dag} $ &  $ \bar B^0 \to K^+  K^-$\\
\hline
\end{tabular}
\end{center}
\caption{$95\%$ C.L. limits on the products of $\lpp$-type
$R$-parity couplings assuming 100 GeV common sfermion mass.
The limits correspond to the label $(*)$ are at $90\%$ C.L., while
the limits denoted by $(\dag)$ are obtained from decay modes with 
only annihilation contributions. }
\label{rpv_b}
\end{table}
\newpage
\begin{table}[ht]
\footnotesize
\begin{center}
\begin{tabular}{|c|c|c|} \hline
Couplings & Bound & {Process} \\
\hline
              & ${1.49\times 10^{-3}}^{*}$& $\bar B^0 \to \pi^0 \pi^0 $\\
\cline{2-3}
 $\mid \lp_{i11}\lp_{i13}\mid $        &  $ 1.86\times 10^{-3} $ & $\bar B^0 \to \pi^+ \pi^- $\\
\cline{2-3}
         &  $ 1.13\times 10^{-2*\dag} $ & $\bar B^0 \to K^- K^0 $\\
\cline{2-3}
      &  $ {3.68\times 10^{-2}}^{*\dag} $ & $\bar B^- \to K^- K^+ $\\
\cline{2-3}
              & ${1.39\times 10^{-2}}^{*\dag}$& $\bar B^0 \to K^0 \bar K^0 $\\
\hline
         & $ {1.95\times 10^{-3}}^{*} $ & $\bar B^0 \to \pi^0 \pi^0$\\
\cline{2-3}
$\mid \lp_{i31}\lp_{i11}\mid $  & $ 2.91\times 10^{-2\dag} $ & $\bar B^0 \to \pi^+ \pi^-$\\
\cline{2-3}
            & $ {3.49 \times 10^{-3}}^{*} $ & $ B^- \to \pi^- \pi^0$\\
\hline
          & $ 3.16\times 10^{-2} $ & $ B^- \to K^- \pi^+$\\
\cline{2-3}
$ \mid \lp_{i12}\lp_{i13}\mid $ & $ 2.78 \times 10^{-3}$ & $ B^- \to K^- \pi^0 $\\
\cline{2-3}
         & $2.66 \times 10^{-3}$ & $ B^- \to \bar K^0 \pi^- $ \\
\cline{2-3}
         & $ 1.71 \times 10^{-3}$ & $ \bar B^0 \to \bar K^0 \pi^0 $ \\
\hline
         & $ 4.19 \times 10^{-2\dag} $ & $ \bar B^0 \to K^- \pi^+ $ \\
\cline{2-3}
$ \mid \lp_{i31}\lp_{i21} \mid $ & $ 1.33 \times 10^{-3}$ & $ B^- \to K^- \pi^0 $ \\
\cline{2-3}
          & $ 2.13 \times 10^{-3} $ & $ B^- \to \bar K^0 \pi^- $ \\
\cline{2-3}
           & $ 1.99 \times 10^{-3} $ & $ \bar B^0 \to \bar K^0 \pi^0 $ \\
\hline
         & $ 2.64 \times 10^{-1\dag} $ & $ \bar B^0 \to K^- \pi^+ $ \\
\cline{2-3}
$ \mid \lp_{i11}\lp_{i23}\mid, \mid \lp_{i32}\lp_{i11}\mid $ & $ 3.28 \times 10^{-2}$ & $ B^- \to K^- \pi^0 $ \\
\cline{2-3}
          & $ 8.30 \times 10^{-3} $ & $ B^- \to \bar K^0 \pi^- $ \\
\cline{2-3}
           & $ 2.18 \times 10^{-3} $ & $ \bar B^0 \to \bar K^0 \pi^0 $ \\
\hline
         & $ {2.88 \times 10^{-3}}^{*} $ & $ B^- \to K^- K^0 $ \\
\cline{2-3}
$ \mid \lp_{i22} \lp_{i13} \mid,\mid \lp_{i31} \lp_{i22} \mid $ & $ {5.26 \times 10^{-3}}^{*} $ & $ \bar B^0 \to K^0 \bar K^0 $ \\
\cline{2-3}
          & $ {5.83 \times 10^{-3}}^{*\dag} $ & $ \bar B^0 \to K^- K^+ $ \\
\hline
$\mid \lp_{i21}\lp_{i23}\mid, \mid \lp_{i32}\lp_{i12}\mid $ & $ {6.70 \times 10^{-4}}^{*} $ & $ B^- \to K^- K^0 $ \\
\cline{2-3}
 & $ {9.49 \times 10^{-4}}^{*} $ & $ \bar B^0 \to K^0 \bar K^0 $ \\
 \cline{2-3}
 & $ {3.68 \times 10^{-2}}^{*\dag} $ & $ \bar B^0 \to K^- \bar K^+ $ \\
\hline
\end{tabular}
\end{center}
\caption{$95\%$ C.L. limits on the products of $\lambda'$ 
$R$-parity violating couplings for a common sfermion mass 
$ m_{\tilde f} = 100 $ GeV.
The limits correspond to the label $(*)$ are at $90\%$ C.L., while the 
limits denoted denoted by $(\dag)$ are from decay modes with only 
annihilation contributions.}
\label{rpv_l}
\end{table}
 \newpage
\begin{table}[ht]
\footnotesize
\begin{center}
\begin{tabular}{|c|c|c|l|} \hline
Product of couplings & our limits & previous limits & process of others
constraints  \\
\hline
$\lambda_{112}^{''}\lambda_{113}^{''}$&$[6.80\times 10^{-3}]$
&$2\times 10^{-8}$&$n-\bar n$ and double neucleon decay\cite{apv}\\
\hline
$\lambda_{212}^{''}\lambda_{213}^{''}$&$1.03\times 10^{-2}$
&$1.5$&perturbativity bound\cite{apv}\\
\hline
$\lambda_{312}^{''}\lambda_{313}^{''}$&$1.03\times 10^{-2}$
&$1.5$&perturbativity bound\cite{apv}\\
\hline
$\lambda_{112}^{''}\lambda_{123}^{''}$&$[3.30\times 10^{-3*}]$
&$2\times 10^{-8}$&double nucleon decay\cite{apv} \\
\hline
$\lambda_{212}^{''}\lambda_{223}^{''}$&$3.19\times 10^{-3*}$
&$1.5$&perturbativity bound\cite{apv}\\
\hline
$\lambda_{312}^{''}\lambda_{323}^{''}$&$3.19\times 10^{-3*}$
&$1.5$&perturbativity bound\cite{apv}\\
\hline
$\lambda_{111}^{'}\lambda_{113}^{'}$&$[1.49\times 10^{-3*}]$
&$1.1\times 10^{-5}$& \\
\hline
$\lambda_{211}^{'}\lambda_{213}^{'}$&$1.49\times 10^{-3*}$    &$3.5\times
10^{-3}$& \\
\hline
$\lambda_{311}^{'}\lambda_{313}^{'}$&$1.49\times 10^{-3*}$    &$3.6\times
10^{-3}$&$\Delta m_{B_d}$\cite{gg1} \\
\hline
$\lambda_{111}^{'}\lambda_{123}^{'}$&$[2.18\times 10^{-3}]$ &$2.2\times
10^{-5}$& \\
\hline
$\lambda_{211}^{'}\lambda_{223}^{'}$&$2.18\times 10^{-3}$ &$1.2\times
10^{-2}$& \\
\hline
$\lambda_{311}^{'}\lambda_{323}^{'}$&$2.18\times 10^{-3}$ &$1.6\times
10^{-2}$&$\Delta
m_{B_d}$\cite{gg1} \\
\hline
$\lambda_{111}^{'}\lambda_{131}^{'}$&$[1.95\times 10^{-3}]$    &$1.0\times
10^{-5}$&\\
\hline
$\lambda_{211}^{'}\lambda_{231}^{'}$&$1.95\times 10^{-3}$    &$1.1\times
10^{-2}$&\\
\hline
$\lambda_{311}^{'}\lambda_{331}^{'}$&$1.95\times 10^{-3}$    &$5.0\times
10^{-2}$&\\
\hline
$\lambda_{111}^{'}\lambda_{132}^{'}$&$[2.18\times 10^{-3}]$
&$1.4\times 10^{-4}$& \\
\hline
$\lambda_{211}^{'}\lambda_{232}^{'}$&$[2.18\times 10^{-3}]$
&$4.7\times 10^{-4}$&$\Delta m_K$\cite{gg1} \\
\hline
$\lambda_{311}^{'}\lambda_{332}^{'}$&$[2.18\times 10^{-3}]$
&$4.7\times 10^{-4}$&$\Delta m_K$\cite{gg1} \\
\hline
$\lambda_{112}^{'}\lambda_{113}^{'}$&$[1.71\times 10^{-3}]$    &$4.4\times
10^{-4}$&\\
\hline
$\lambda_{212}^{'}\lambda_{213}^{'}$&$1.71\times 10^{-3}$    &$3.5\times
10^{-3}$&\\
\hline
$\lambda_{312}^{'}\lambda_{313}^{'}$&$1.71\times 10^{-3}$    &$1.2\times
10^{-2}$&\\
\hline
$\lambda_{112}^{'}\lambda_{132}^{'}$&$6.70\times 10^{-4*}$&$5.9\times
10^{-3}$& \\
\hline
$\lambda_{212}^{'}\lambda_{232}^{'}$&$6.70\times 10^{-4*}$&$3.3\times
10^{-2}$& \\
\hline
$\lambda_{312}^{'}\lambda_{332}^{'}$&$6.70\times 10^{-4*}$&$5.0\times
10^{-2}$& \\
\hline
$\lambda_{113}^{'}\lambda_{122}^{'}$&$[2.88\times 10^{-3*}]$&$9.0\times
10^{-4}$&\\
\hline
$\lambda_{213}^{'}\lambda_{222}^{'}$&$2.88\times 10^{-3*}$&$1.2\times
10^{-2}$&\\
\hline
$\lambda_{313}^{'}\lambda_{322}^{'}$&$2.88\times 10^{-3*}$&$5.7\times
10^{-2}$&\\
\hline
$\lambda_{121}^{'}\lambda_{123}^{'}$&$6.70\times 10^{-4*}$&$1.4\times
10^{-3}$&$\Delta
m_{B_d}$\cite{gg1}\\
\hline
$\lambda_{221}^{'}\lambda_{223}^{'}$&$6.70\times 10^{-4*}$&$1.4\times
10^{-3}$&$\Delta
m_{B_d}$\cite{gg1}\\
\hline
$\lambda_{321}^{'}\lambda_{323}^{'}$&$6.70\times 10^{-4*}$&$1.4\times
10^{-3}$&$\Delta
m_{B_d}$\cite{gg1}\\
\hline
$\lambda_{121}^{'}\lambda_{131}^{'}$&$[1.33\times 10^{-3}]$    &$8.2\times
10^{-4}$&\\
\hline
$\lambda_{221}^{'}\lambda_{231}^{'}$&$1.33\times 10^{-3}$    &$3.2\times
10^{-2}$&\\
\hline
$\lambda_{321}^{'}\lambda_{331}^{'}$&$1.33\times 10^{-3}$    &$2.3\times
10^{-1}$&\\
\hline
$\lambda_{122}^{'}\lambda_{131}^{'}$&$[2.88\times 10^{-3*}]$&$1.0\times
10^{-4}$&$\Delta
m_K$\cite{gg1}\\
\hline
$\lambda_{222}^{'}\lambda_{231}^{'}$&$[2.88\times 10^{-3*}]$&$1.0\times
10^{-4}$&$\Delta
m_K$\cite{gg1}\\
\hline
$\lambda_{322}^{'}\lambda_{331}^{'}$&$[2.88\times 10^{-3*}]$&$1.0\times
10^{-4}$&$\Delta
m_K$\cite{gg1}\\
\hline
\end{tabular}
\end{center}
\caption{
Comparison of constraints obtained in this paper with other
existing constraints.
$95\%$ C.L. limits on the products of different $R$-parity
couplings for sfermion masses $m_{\tilde f} = 100 $ GeV.
The limits correspond to the label $(*)$ are at $90\%$ C.L.. The limits
shown in the square bracket are weaker than the existing bounds. The
bounds
corresponding to the processes not shown in the Table are coming from the
product of individual couplings~\cite{apv}.
}
\label{rpv_tot}
\end{table}
\newpage
\begin{figure}[hbt]
\centerline{\epsfig{file=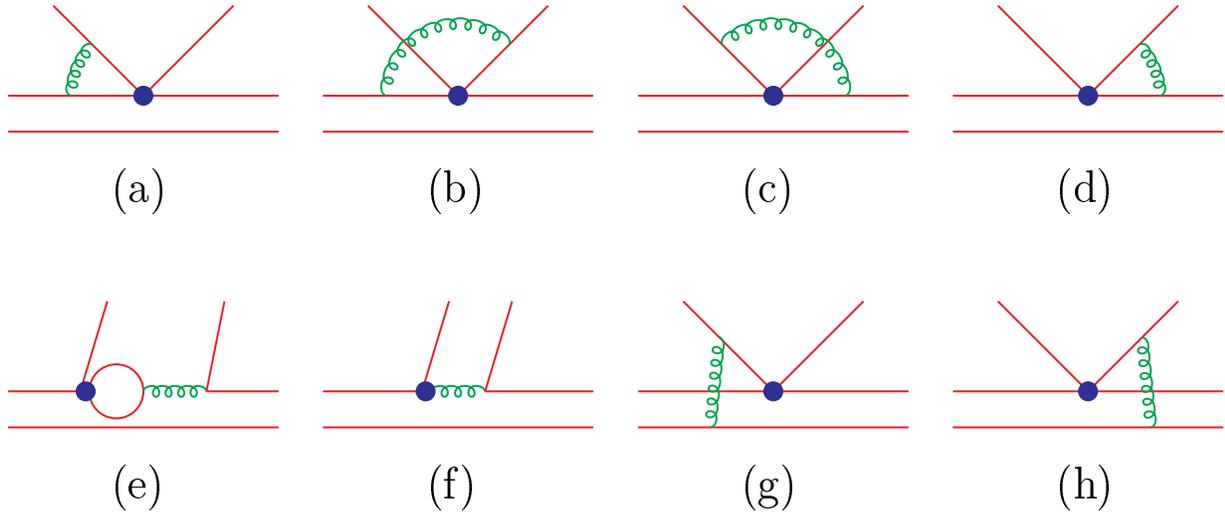,width=\linewidth}}
\vspace{1.0in}
\caption{\it Order of $\alpha_s$ corrections to the hard-scattering kernels
$T^I $ and $T^{II}$. The quark lines directed upwards represent the ejected
quark pairs from weak decays of $b$-quark.}
\label{fig:kern}
\end{figure}
\begin{figure}[hbt]
\centerline{\epsfig{file=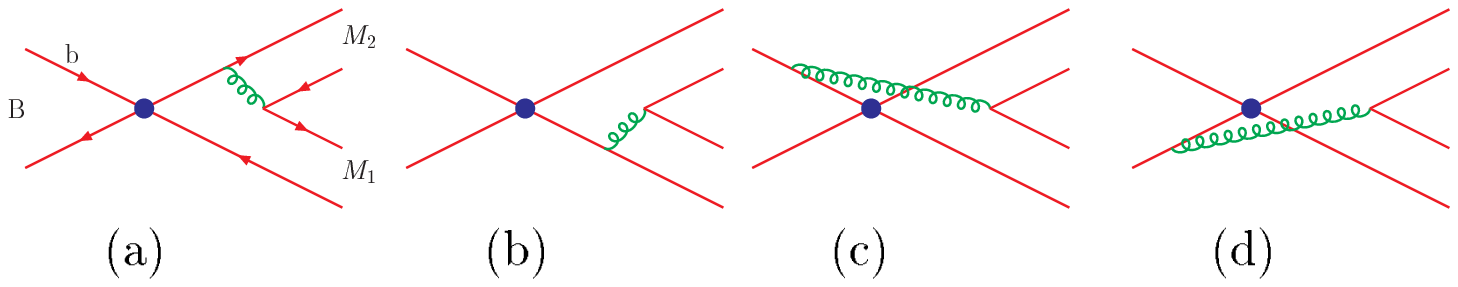,width=\linewidth}}
\caption{\it Annihilation diagrams}
\label{fig:annih}
\end{figure}

\begin{thebibliography}{99}
\bibitem{susy}
H.~P.~Nilles, \prp110, 1 {84};
H.~E.~Haber and G.~L.~Kane, \prp117, 75 {85}.

\bibitem{BBNS0}M.~Beneke, G.~Buchalla, M.~Neubert and C.~T.~Sachrajda,
\prl83, 1914 {99}.

\bibitem{hn} Y.-Y. Keum, H.-n. Li and I. Sanda, Phys. Lett. {\bf B504},
6(2001); Phys. Rev. {\bf D63}, 054008(2001).

\bibitem{BBNS1}M.~Beneke, G.~Buchalla, M.~Neubert and C.~T.~Sachrajda,
\pnpb591, 313 {00}.


\bibitem{DDG}D.~S.~Du, H.~-U.~Gong, J.-F.~Sun, D.~-S.~Yang  and
G.~-H.~Zhu, hep-ph/0108141;
D.~-S.~Du, D.~-S.~Yang and G.~-H.~Zhu, \pprd64, 014036 {01}; 
\pplb509, 263 {01}.

\bibitem{he1} J.~Chay, \pplb476, 339 {00};
T.~Muta, A.~Sugamoto, M.~Z.~Yang and Y.~D.~Yang, \pprd62, 094020 {00};
D.~S.~Du, D~.S.~Yang and G.~H.~Zhu, \pplb488, 46 {00};
M.~Z.~Yang and Y.~D.~Yang, \pprd62, 114019 {00};
J.~Chay and C.~Kim, hep-ph/0009244;
X.~-G.~ He, J.~-P.~ Ma and C.~-Y.~ Wu,\pprd63, 094004 {01};
H.~-Y.~Cheng and K.~-C.~Yang, \pplb511, 40 {01}; \pprd63, 074011 {01};
M. Chiuchini, E.~Franco, G.~Martinelli, M.~Pierini and 
L.~Silvestrini,  hep-ph/0110022.

\bibitem{he2} X.-G. He, J.-Y. Lieu and J.-Q. Shi, \pprd64, 094018 {01}.

\bibitem{proton_dk} Particle Data Group, D.~E.~Groom \etal, \eepj15, 1 {00}.

\bibitem{fayet}
P.~Fayet, \plb69, 489 {77};
G.~Farrar and P.~Fayet, \plb76, 575 {78};
N.~Sakai and T.~Yanagida, \npb197, 533 {82};
C.~Aulakh and R.~Mohapatra, \plb119, 136 {83}.

\bibitem{weinberg} S.~Weinberg, \prd26, 287 {82}.

\bibitem{rpv_review}
\label{rpv_review}
For recent reviews on $R$-parity violation see
G.~Bhattacharyya,~hep-ph/9709395 and references therein;
H.~Dreiner, {\it An Introduction to Explicit R-Parity Violation} in
{\it Perspectives on Supersymmetry}, p.462-479, Ed. G.L.~Kane
(World Scientific) and references therein;
R.~Barbier {\it et al}, {\it Report of the Group on the $R$-parity Violation},
hep-ph/9810232 (unpublished);
B.~C.~Allanach, A.~Dedes and H.~K.~Dreiner,\prd60, 075014 {99} and references 
therein. 

\bibitem{Majorana}
\label{Majorana}
L.~Hall and M.~Suzuki, \npb231, 419 {84};
R.~Barbirei and A.~Masiero, \npb267, 679 {86};
R.~Mohapatra, \prd34, 3457 {86};
K.~Babu and R.~Mohapatra, \prl75, 2276 {95};
M.~Hirsch, H.~Klapdor-Kleingrothaus and S.~Kovaleno, \prl75, 17 {95}; \prd53, 
1329 {96};
A. Wodecki, W. kominski and S. Pagerka, Phys. Lett. {\bf B413}, 342(1997).

\bibitem{han}
V.~Barger, G.~F.~ Giudice and T.~Han, \prd40, 2987 {89};
D. Lebedev, W. Loinaz and T. Takeuchi, Phys. Rev. {\bf D61}, 115005(2000);

\bibitem{proton}
F.~Zwirner, \plb132, 103 {83};
J.~Goity and M.~Sher, \plb346, 69 {95};
C.~Carlson, P.~Roy and M.~Sher, \plb357, 99 {95}; 
I.~Hinchliffe and T.~Kaeding, \prd47, 279 {93};
A.~Y.~Smirnov and F.~Vissani, \plb380, 317 {96};
G. Bhattachayya and P. B. Pal, Phys. Lett. {\bf B439}, 81(1998).


\bibitem{Zdk}
R.~Godbole, P.~Roy and X.~Tata, \npb401, 67 {93};
G.~Bhattacharyya, J.~Ellis and K.~Sridhar, \mpl10, 1583 {95};
G.~Bhattacharyya and D.~Choudhury, and K.~Sridhar, \plb355, 193 {95};
K.~Agashe and M.~Graesser, \prd54, 4445, {96};
R.~Barbieri, A.~Strumia and Z.~Berezhiani, \plb407, 250 {97}; 
D. Choudhury and S. Raychaudhury, Phys. Lett. {\bf 401}, 54(1997).

\bibitem{dc1} D. Chakraverty and D.~Choudhury, \pprd63, 112002 {01};
D.~Chakraverty and D.~Choudhury, \pprd63, 075009 {01};
G.~Bhattacharyya, A.~Datta and A.~Kundu, \pplb514, 47 {01}; 
D.~E.~Kaplan, hep-ph/9703347; S.~A.~Abel, \plb410, 173 {97};
B.~de~Carlos and P.~White, \prd55, 4222 {97};
G.~Bhattacharyya and D.~Choudhury, \mpl10, 1699 {95}; 

\bibitem{gamma}
K. Enqvist, A. Masiero and A. Riotto, Nucl. Phys. {\bf B373}, 95(1992);
D.~Choudhury and P.~Roy, \plb378, 153 {96};
M.~Chaichian and K.~Huitu, \plb384, 157 {96};
K.~Huitu, M.~Raidal, and A.~Santamaria \plb430, 355 {98};
K. Cheung and R.-J. Zhang, \plb427, 73 {98};
M. Frank, plb463, 234 {99}.

\bibitem{btolep1}
\label{cpv}
D.~Guetta and E.~Nardi, \prd58, 012001 {98}; Y.~Grossman, Z.~Ligeti
 and E.~Nardi, \prd55, 2768 {97};
J-H.~Jang, J.~Kim and J.~Lee, \prd55, 7296 {97};
J-H. Jang, Y.~G.~Kim and
J.~S.~ Lee, \plb408, 367 {97};
S. Baek and Y.G. Kim, Phys. rev. {\bf D60}, 077701(1999); 
J-H. Jang, Y.~G.~Kim and J.~S.~Lee, \prd58, 035006 {98}.


\bibitem{apv} For updates limits, see B.~C.~Allanach \etal
in \cite{rpv_review}.

\bibitem{gg1} 
\label{aa}
G.~Bhattacharyya, and A.~Raychaudhuri, \prd57, 3837 {98};


\bibitem{lcda}V.~M.~Braun and I.~E.~Filyanov, \zpc48, 239 {90}.

\bibitem{ball}P.~Ball, \jhep9901, 010 {99}. 

\bibitem{form_fac}A.~Abada, D.~Becirevic, P.~Boucaud, J.~P.~Leroy , V.~Lubicz,
G.~Martinelli and F.~Mescia,
{\it Nucl.~Phys.~Proc.~Suppl.~}{\bf 83}, 268 (2000).
  
\bibitem{he3}X.~-G.~ He, Y.~-K.~Hsiao, J.~-Q.~Shi, Y.~-L.~Wu and 
Y.~-F.~Zhou  \pprd64, 034001 {01}. 

\bibitem{Bexp}  D. Cronin-Hennessy, et al. (CLEO collaboration),
\pprl85, 515 {00}; K. Abe, \etal (Belle Collaboration),
\pprl87, 101801 {01};
B. Aubert, \etal (BABAR COllaboration), \pprl87, 151802 {01}. 
\end{thebibliography}
\end{document}